\documentclass[prd,onecolumn,superscriptaddress,nofootinbib,12pt]{revtex4-2}
\usepackage{graphicx}
\usepackage{epsfig}
\usepackage{bm}
\usepackage{amsfonts}
\usepackage{subfigure}
\usepackage{multirow}
\usepackage{float}
\usepackage{color}
\usepackage[usenames,dvipsnames]{xcolor}
\usepackage{ulem}
\usepackage{todonotes}
\usepackage{makecell}
\usepackage{enumitem}
\usepackage{amsmath}
\usepackage{amssymb}
\usepackage{caption}
\usepackage{verbatim}
\usepackage{array}
\usepackage[colorlinks=true,citecolor=Red,linkcolor=Green,urlcolor=Green]{hyperref}

\renewcommand{\arraystretch}{1.5}

\def\beq{\begin{equation}}
	\def\eeq{\end{equation}}
\def\ber{\begin{eqnarray}}
	\def\eer{\end{eqnarray}}
\def\benu{\begin{enumerate}}
	\def\eenu{\end{enumerate}}

\def\sq{\lower.25ex\hbox{\large$\Box$}}
\def \lleq {\lower0.9ex\hbox{ $\buildrel < \over \sim$} ~}
\def \ggeq {\lower0.9ex\hbox{ $\buildrel > \over \sim$} ~}

\usepackage{geometry}
\begin{document}

	\title{\textbf{Radial kinks in the boson stars }}
	\author{Tian-Chi Ma}\email{tianchima@buaa.edu.cn}
	\affiliation{Center for Gravitational Physics, Department of Space Science, Beihang University, Beijing 100191, China}
	\author{Xiang-Yu Wang}\email{xiangyuwang@buaa.edu.cn}
	\affiliation{Center for Gravitational Physics, Department of Space Science, Beihang University, Beijing 100191, China}
	\author{Hai-Qing Zhang}\email{hqzhang@buaa.edu.cn}
	\affiliation{Center for Gravitational Physics, Department of Space Science, Beihang University, Beijing 100191, China}
	\affiliation{Peng Huanwu Collaborative Center for Research and Education, Beihang University, Beijing 100191, China}

	\begin{abstract}
		In this work, we study the time evolution of radial kinks in the background of boson stars. In particular, we consider two types of boson stars: the massive boson star and the solitonic boson star. For each boson star, we study the dynamics of the kinks with four different compactnesses. We observe that the greater the compactness is, the slower the kinks move towards the origin of the boson stars, indicating that the compactness will hinder the kinks to collide with the origin. Additionally, it is found that when the boson star is highly compact, a new kink may turn out after the kink colliding with the origin, instead of immediately dissipating into the background. From this,  we then propose that the radial kinks may potentially serve as a means to probe the internal structures of dense astrophysical objects, even the interior structure of black holes.
		
	\end{abstract}
	
	\maketitle

\section{Introduction}

Topological defects are supposed to be generated in the early stages of the universe \cite{kibble1982phase, vilenkin1994cosmic, vachaspati2006kinks}. According to the types of the symmetry breaking, topological defects can be categorized into various kinds, such as domain walls, vortices, monopoles and etc. \cite{pismen1999vortices,manton2004topological,bunkov2000topological}.  Among these, domain walls (also commonly referred to as kinks), may be the most well-known type of topological defect. Domain walls are related to the spontaneous breaking of $Z_2$ symmetry, which divides the ground states into two distinct vacuum states. Fundamental constants, such as particle mass, fine-structure constant and etc., may change after passing through a domain wall \cite{kibble1976topology,vilenkin1982cosmic,vilenkin1985cosmic}. Domain walls are often considered as possible candidates for the dark matter or other interesting structures \cite{press1989dynamical,avelino2008dynamics,friedland2003domain}. Since domain walls can alter fundamental physical constants, scientists have developed various experiments to detect their existence, including satellite synchronization \cite{derevianko2014hunting,roberts2017search,kalaydzhyan2017extracting}, electric dipole moment measurements \cite{stadnik2014searching}, magnetometer measurements \cite{pustelny2013global,afach2021search}, mass difference causing acceleration \cite{mcnally2020constraining} and gravitational wave detectors \cite{hall2018laser,grote2019novel,jaeckel2021probing}. Therefore, studying the domain walls or kinks in the gravitational system is a very important task.

The evolution of domain walls in Minkowskian spacetime has been extensively studied, including in $2+1$ and $3+1$-dimensional scenarios. Ref. \cite{kevrekidis2018planar} investigates the dynamical evolution of planar and radial kinks in the Klein-Gordon models, exploring their existence and stability. The attractive interaction between kink and antikink has been studied in Ref. \cite{carretero2022kink}, where the authors derived an effective model using the variational approximation method, which is consistent with the results of the full dynamical model. The authors in \cite{caputo2013radial} investigate the effect of radial perturbations on the kink and explores the possibility of extracting energy from the kink. Recently, in anti-de Sitter spacetime, some studies have employed the gauge/gravity duality \cite{maldacena1999large,witten1998anti} to examine the statistical behavior of kinks in the strongly coupled regime \cite{li2023black,ma2025universal}. More related studies can be found in references \cite{arodz1998expansion,dobrowolski2008construction,dobrowolski2009kink,gatlik2021modeling,gorria2004kink,shi2024topologically}. 

In this context, an interesting question arises: how do topological defects, such as domain walls, evolve in the spacetime background of massive celestial bodies --- black holes or stars? Such kind of studies are relatively rare except for some few examples. In the early stages, studies concentrated on the steady-state behavior of thick domain walls in the background of charged dilaton black holes. And, it was found that extremal dilaton black holes exhibit a repulsive behavior towards the domain wall \cite{moderski2003thick}. Later, related studies were extended to Schwarzschild black hole, Reissner-Nordstr\"om black hole and AdS black hole spacetimes \cite{morisawa2003thick,moderski2004reissner,moderski2006thick}. Additionally, some researchers explored the evolution of planar kinks in the Schwarzschild and the Kerr black holes, observing a behavior resembling the ringing modes \cite{ficek2018planar}. However, this line of research has seen little follow-up in the following years. Only recently, researchers have examined the dynamical evolution of radial kinks in the neutron star spacetimes, finding that the presence of a neutron star can slow down the collapse of the kinks towards the center \cite{caputo2024radial}. 

It is well-known that classical general relativity inevitably introduces singularities into black holes, where the physical laws, including general relativity itself, may break down \cite{hawking1970singularities, penrose1965gravitational, senovilla1998singularity}. This highlights the limitations of general relativity. Scientists generally believe that the singularity problem will ultimately be resolved by quantum gravity. However, there is currently no widely accepted theory of quantum gravity. As a result, some scientists have attempted to modify gravitational theories to bypass the singularity problem in black holes, leading to a series of regular black hole solutions, such as the Bardeen black hole \cite{bardeen1968non} and Hayward black hole \cite{hayward2006formation}, and others \cite{ayon1998regular,berej2006regular}. Additionally, scientists have also tried to find analogues of black holes to avoid central singularities, such as boson stars and Proca stars \cite{colpi1986boson,lee1987soliton,vincent2016imaging,ma2023hybrid,ma2025boson,brito2016proca,landea2016charged}. Boson stars are a class of compact objects formed by self-interacting scalar fields, which can achieve a compactness comparable to that of black holes and exhibit some similar observational features, such as photon rings \cite{rosa2023imaging,johnson2020universal,olmo2023shadows,pitz2023generating,he2025observation,zeng2025optical,zeng2025polarization}. Moreover, since the only scalar particle in the Standard Model is the Higgs bosons, and boson stars are clearly not composed of Higgs bosons, studying boson stars may provide novel insights into new physics. 

In this paper, we focus on the evolution of radial kinks in the boson star backgrounds, in particular in the background of spherical massive boson stars and solitonic boson stars. Compared to the case in Minkowskian spacetime, the scalar fields in the boson stars will hinder the kinks to move towards the center or origin of the gravity background. Besides, we find that the compactness of the boson stars will affect the kink dynamics, i.e., the larger compactness will more delay the kinks to move towards the origin. After the collision of the kinks with the origin, the scalar fields will fluctuate significantly and then become ripples and dissipate into the background. However, for both boson stars with larger compactness, after the collision we observe the new formed transient kinks, although they finally dissipate into the background as well. In the solitonic boson star with very high compactness, after the collision, we can even observe a new formed kink moving much far away from the origin. This indicates that the process of the kink entering the interior of the boson star and subsequently moving outward might carry some information from within the boson star. Therefore, it can potentially provide some clues for the internal structures of the dense celestial objects. This is a very strong motivation for our work --- the kink dynamics may extract information from the interior of the boson stars. 

This paper is arranged as follows. In Sec. \ref{sect2}, we present the theoretical framework of the boson stars and the radial kinks. In Sec. \ref{sect4}, we investigate the evolution of the radial kink in the background of boson stars, in particular in the massive boson stars and solitonic boson stars. Finally, we draw conclusions and discussions in Sec. \ref{sect5}. Additionally, we review the evolution of the radial kinks in the Minkowskian spacetime in Appendix \ref{sect3}.

\section{Theoretical framework}\label{sect2}

\subsection{Boson star models and configurations}
In order to get the background of boson star, we consider the following action
\begin{equation} \label{eq:action}
	\mathcal{S}=\int d^4x\sqrt{-g}\left[\frac{R}{16\pi}-{\nabla}_a\Phi^*{\nabla}^a\Phi-U(|\Phi|^2)\right],
\end{equation}
where $R$ is the Ricci scalar, $\Phi$ is the complex scalar field and $U$ is the scalar potential. Varying Eq. (\ref{eq:action}) with respect to the metric $g_{\mu\nu}$ and $\Phi$ yields the Einstein equation and the equation for the scalar field respectively, 
\begin{align}
	G_{ab}-8\pi T_{ab}=0,\label{eq:einstein}\\
	\nabla_a\nabla^a\Phi - \dot{U}({|\Phi|^2}) \Phi=0,\label{eq:scalar}
\end{align}
where $G_{ab}$ is the Einstein tensor, $\dot{U}$ represents the derivative of $U$ with respect to $|\Phi|^2$, and $T_{ab}$ is the energy-momentum tensor of the complex scalar field $\Phi$ given by
\begin{equation}
	T_{ab}=\nabla_a\Phi^*\nabla_b\Phi+\nabla_b\Phi^*\nabla_a\Phi-g_{ab}(\nabla_c\Phi^*\nabla^c\Phi+U).
\end{equation}
For a static and spherically symmetric spacetime we consider the following ansatz for the line element and scalar field, 
\begin{align}
	ds^2&=-A(r) dt^2+B(r)^{-1}dr^2+r^2d\Omega^2,\\
	\Phi&=\phi(r) e^{i \omega t},\label{eq:Phi}
\end{align}
where the metric functions $A(r)$ and $B(r)$ only depends on the radial direction, $\phi(r)$ describes the radial dependence of the scalar field with $\omega$ representing its frequency, and $d\Omega^2=d\theta^2+\sin^2\theta d\varphi^2$ with $\theta$ and $\varphi$ the polar angle and azimuthal angle, respectively.   By substituting this ansatz into Eqs. \eqref{eq:einstein} and \eqref{eq:scalar}, we obtain the following equations for the system,
\begin{align}
	&A'-\frac{A(1-B)}{Br}-8\pi r\left(\frac{\omega^2 \phi^2}{B}+A\phi'^2-\frac{A}{B}U\right)=0,\label{eq:EQofMo1}\\
	&B'-\frac{1-B}{r}+8\pi r\left(\frac{\omega^2 \phi^2}{A}+B\phi'^2+U\right)=0,\label{eq:EQofMo2}\\
	&\phi''+\left(\frac{2}{r}+\frac{1}{2}\left(\frac{A'}{A}+\frac{B'}{B}\right)\right)\phi'+\frac{1}{B}\left(\frac{\omega^2}{A}-\frac{dU}{d\phi^2}\right)\phi=0,\label{eq:EQofMo3}
\end{align}
where ``$\ '\ $'' denotes the differentiation with respect to $r$. These equations are highly nonlinear, and solving them requires appropriate numerical techniques. We will employ the shooting method to solve the equations, following the procedures outlined in Reference \cite{rosa2023imaging}. To achieve this goal, we impose the boundary conditions by enforcing the asymptotic flatness of the geometry, similar to the Schwarzschild solution, and require the radial scalar field to vanish as $r \to \infty$. That is
\begin{align}
	A(r\to\infty)=A_\infty\left(1-\frac{2M}{r}\right),\ \ \ \ \ \ 
	B(r\to\infty)=1-\frac{2M}{r},\ \ \ \ \ \ 
	\phi(r\to \infty)=0,\label{eq:BCatInf}
\end{align}
where $M$ denotes the total mass of the boson star. Near the origin $r\approx0$ \footnote{Since Eqs. \eqref{eq:EQofMo1}-\eqref{eq:EQofMo3} include terms proportional to $1/r$, which diverge at the origin. Therefore, we begin the numerical integration at a point very close to the origin.}, we require the metric functions to be finite or normalized, and set the radial scalar field to be a given central value $\phi_c$, i.e., 
\begin{align}
	A(r\approx 0)=A_c,\ \ \ \ \ \
	B(r\approx 0)=1,\ \ \ \ \ \
	\phi(r\approx 0)=\phi_c,\ \ \ \ \ \ \phi'(r\approx0)=0.\label{eq:BCatOri}
\end{align}
In fact, $A_c$ and $A_\infty$ are not independent: once $A_c$ is given, $A_\infty$ will be automatically determined. Since Eqs. \eqref{eq:EQofMo1}-\eqref{eq:EQofMo3} are time-independent, for any arbitrary constant $a$ the equations remain invariant under the transformation
\begin{align}
	t\to at,\ \ \ \ \ \
	A(r)\to a^{-2}A(r),\ \ \ \ \ \
	\omega\to a^{-1}\omega.\label{}
\end{align}
Therefore, we can always perform a transformation $A(r)\to A(r)/A_\infty$ and $\omega\to\omega/\sqrt{A_\infty}$ to set $A_\infty=1$, which allows us to obtain the spacetime solution in the usual Schwarzschild-like coordinates.  With these boundary conditions, we can numerically integrate the Eqs. \eqref{eq:EQofMo1}-\eqref{eq:EQofMo3} from the origin to the infinite boundary.

In order to investigate the motion of a kink in the background of the boson star, we choose the following two scalar potential functions, 
\ber U=\mu^2|\Phi|^2+\Lambda|\Phi|^4~~~~{\rm and} ~~~~U=\mu^2|\Phi|^2(1-|\Phi|^2/\alpha^2)^2, \eer 
which correspond to the massive boson star \cite{colpi1986boson} and the solitonic boson star \cite{lee1987soliton}, respectively. Here, $\mu$ is the mass term, $\Lambda$ and $\alpha$ are respectively the coupling constants. In order to obtain a more compact boson star configuration, without the loss of generality, we choose the parameters $\mu = 1$, $\Lambda = 400$, and $\alpha =0.08$ throughout this paper. As a result, Eqs. \eqref{eq:EQofMo1}-\eqref{eq:EQofMo3} contain only the frequency $\omega$ as a free parameter, which we vary to generate a series of boson star configurations with different values of compactness.  The dynamical stability of these two types of boson stars has been studied in depth in the literatures \cite{jetzer1992boson,gleiser1988stability,seidel1990dynamical,kusmartsev1991gravitational,brito2023stability,siemonsen2021stability}. According to these references, the boson star configurations considered in our paper are dynamically stable in the selected range of parameters. 

\begin{figure}[]
	\centering
	\includegraphics[trim=3.2cm 9.5cm 3cm 9.5cm, clip=true, scale=0.39]{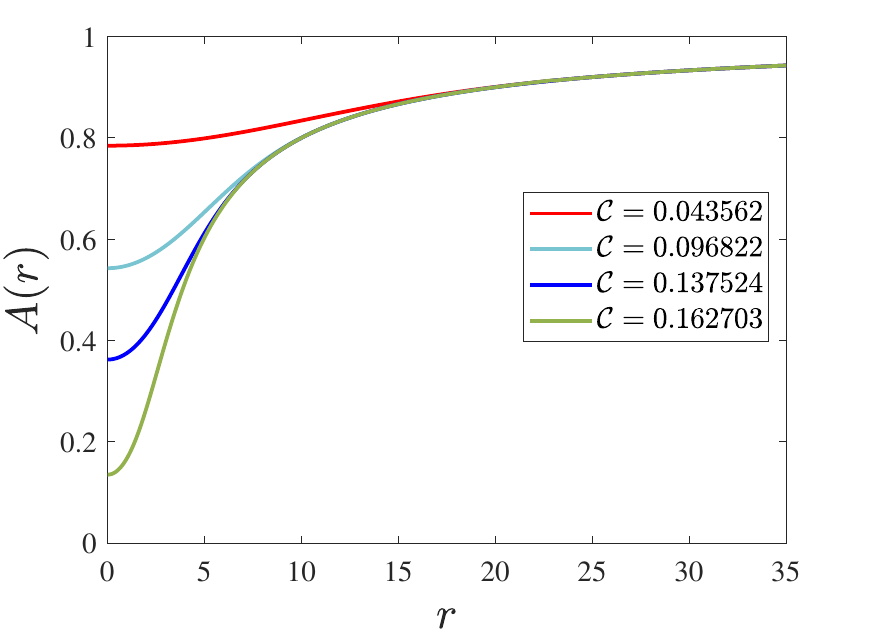}
	\put(-167,108){(a)}
	\includegraphics[trim=3.2cm 9.5cm 3cm 9.5cm, clip=true, scale=0.39]{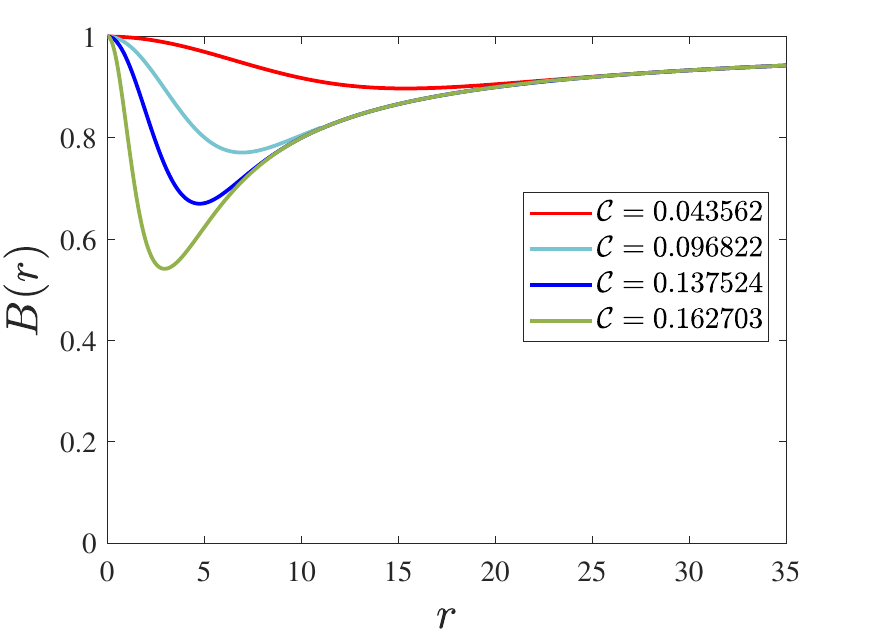}
	\put(-167,108){(b)}
	\includegraphics[trim=3.2cm 9.5cm 3cm 9.5cm, clip=true, scale=0.39]{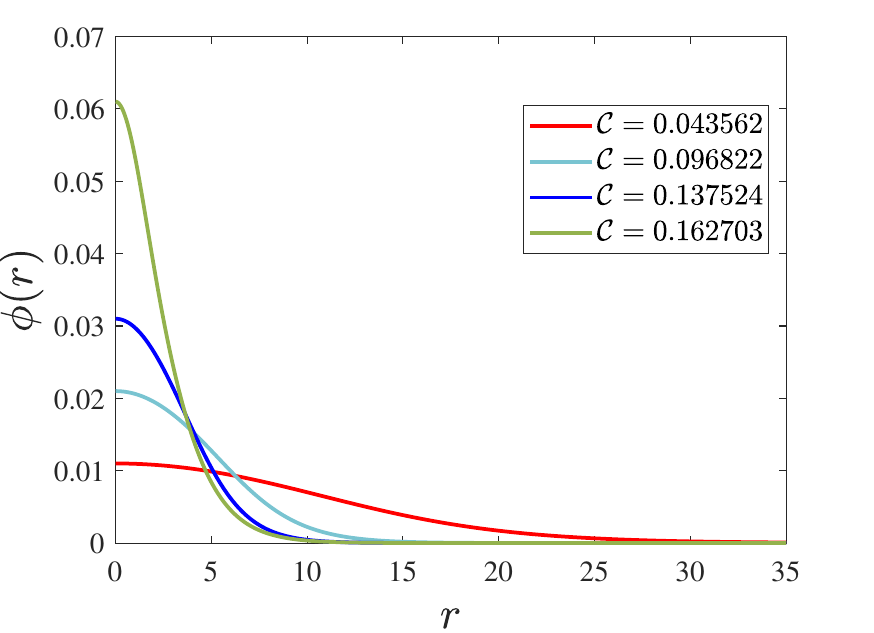}
	\put(-167,108){(c)}
	\caption{Background configurations $A(r)$, $B(r)$ and the scalar field $\phi(r)$ in the massive boson star with four different compactnesses $\mathcal{C}$ = 0.043562, 0.096822, 0.137524 and 0.162703.
	}\label{fig:massive}
\end{figure}

In this work, a spherical initial kink is assumed to start from rest to falling into the center of the boson star from the exterior region. To ensure that the effective radius of the boson star lies within the radial range of the radial kink, we also require the compactness of the boson star to span a relatively large range. Taking into account of the numerical stability of the shooting methods, we select four solutions which are presented in Table \ref{tab:massive} for massive boson stars.  In Fig. \ref{fig:massive}, we present the corresponding field configurations $A(r), B(r)$ and $\phi(r)$ for massive boson star with various values of the compactness $\mathcal{C}$. \footnote{For the parameter choices considered in this paper, the range of the compactness is $\mathcal{C}\in [0, 0.173]$ of the massive boson star.} From the Fig. \ref{fig:massive}(c), we see that the scalar field $\phi(r)$ will ``condensate" near the center $r=0$. Moreover, as the compactness parameter $\mathcal{C}$ increases, the scalar field will more condensate near the center,  which reflects the meaning of the compactness $\mathcal{C}$.

\begin{table*}[]
	\setlength{\tabcolsep}{10pt}
	\renewcommand{\arraystretch}{1}
	\centering
	\begin{tabular}{c c c c c c c}
		\hline\hline
		$\omega/\mu$ & $\phi_c$ &$A_c$ & $\mu M$ & $\mu R$ & $\mathcal{C}$ \\
		\hline\hline
		0.944572 & 0.011 & 0.784557 & 0.790508 & 18.1468 & 0.043562 \\  
		0.878480 & 0.021 & 0.542684 & 1.246500 & 12.8741 & 0.096822\\
		0.828552 & 0.031 & 0.362296 & 1.368091 & 9.94803 & 0.137524\\
		0.797111 & 0.061 & 0.135073 & 1.05039  & 6.45591 & 0.162703\\
		\hline\hline
	\end{tabular}
	\caption{Four different configurations of massive boson star used in this paper. Parameter $\omega$ is the scalar field frequency, $\phi_c$ is the value of scalar field at the origin, $A_c$ is the value of the metric field at the origin, $\mu$ is the mass of the scalar field, $R$ is the radius of the star, $M$ is the total mass of the massive boson star, and $\mathcal{C} \equiv M/R$ is the compactness. }
	\flushleft
	\label{tab:massive}
\end{table*}

\begin{figure}[]
	\centering
	\includegraphics[trim=3.2cm 9.5cm 3cm 9.5cm, clip=true, scale=0.39]{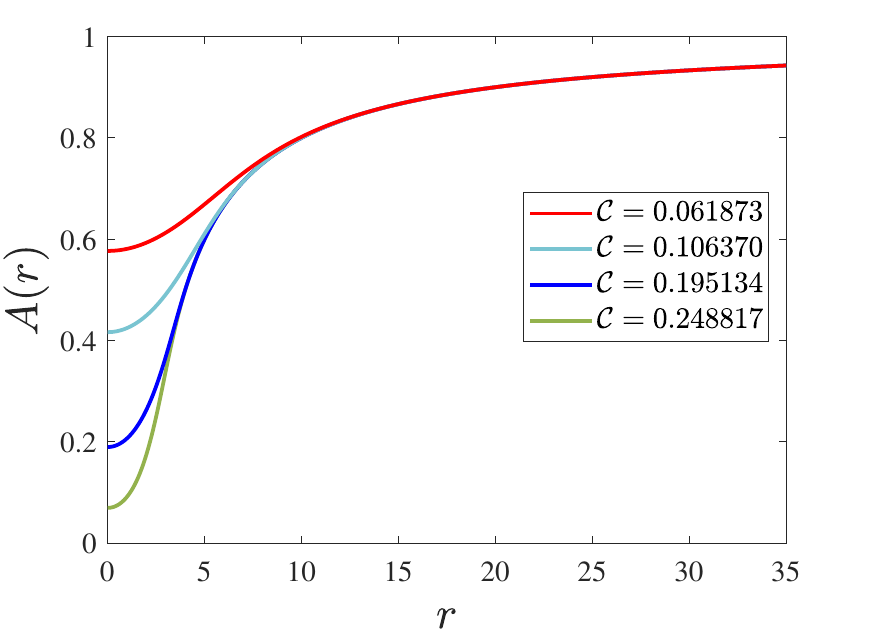}
	\put(-167,108){(a)}~
	\includegraphics[trim=3.2cm 9.5cm 3cm 9.5cm, clip=true, scale=0.39]{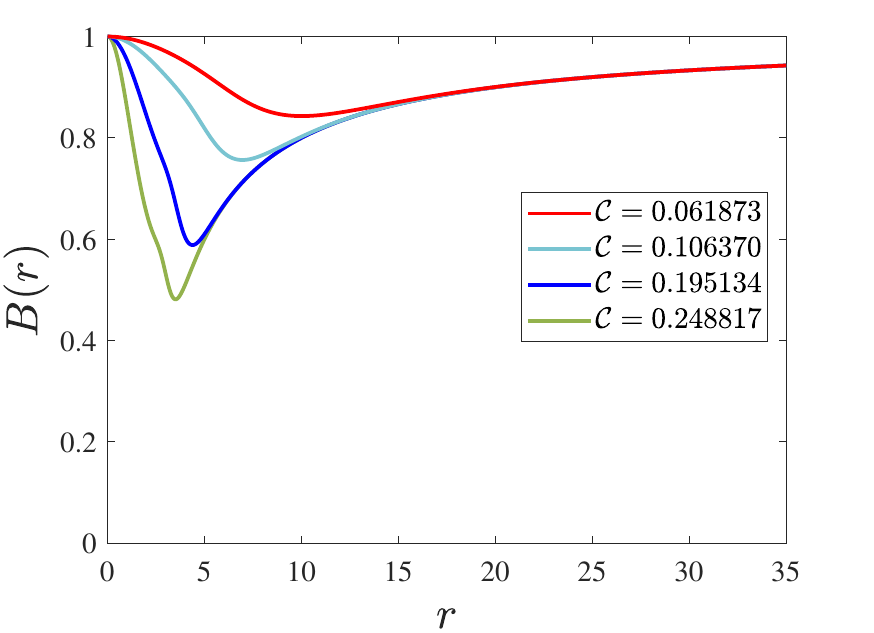}
	\put(-167,108){(b)}~
	\includegraphics[trim=3.2cm 9.5cm 3cm 9.5cm, clip=true, scale=0.39]{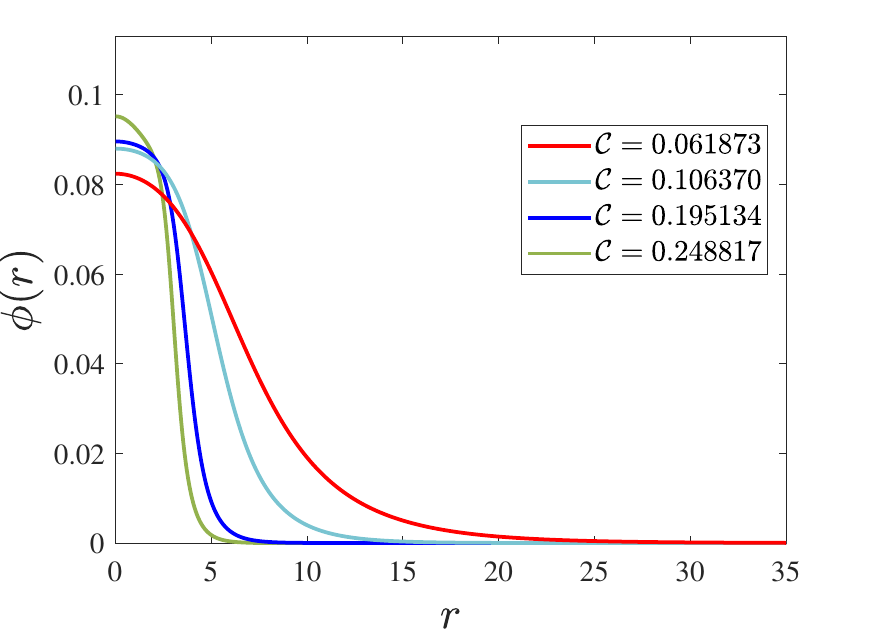}
	\put(-167,108){(c)}
	\caption{Background quantities for the metric coefficients $A(r)$, $B(r)$ and the scalar field $\phi(r)$ of solitonic boson stars with four different compactnesses $\mathcal{C}$ = 0.061873, 0.106370, 0.195134 and 0.248817.
	}\label{fig:soliton}
\end{figure}

Similarly, Table \ref{tab:soliton} and Fig. \ref{fig:soliton} show our parameter choices for the solitonic boson stars and their corresponding field configurations. \footnote{The solitonic boson star is named after its potential, which is one of the simplest potentials exhibiting non-topological soliton characteristics in the absence of gravity \cite{lee1987soliton,lee1992nontopological}. In certain ranges of parameters, a solitonic boson star can have a more compact configuration than a massive boson star. In the limit of $\alpha \to 0$, ultra-compact solutions can be obtained, with the maximum compactness reaching $\mathcal{C} \approx 0.36$ \cite{rosa2023imaging,cardoso2022eco}.} For both of these boson stars, the scalar field $\phi(r)$ exhibits an exponential decay for $r\gg \mu^{-1}$, but it never reaches zero.  Therefore, we define the effective radius of the boson star as the radius that contains $98\%$ of the boson star's total mass. The mass function can be obtained from $B(r)=1-2m(r)/r$, and thus the effective radius is defined as $m(R)=0.98M$. Additionally, the Arnowitt-Deser-Misner (ADM) mass is defined as the asymptotic value of the solution $m(r)$ as \( r \to \infty \), with \( m(\infty) \equiv M \). It is worth noting that in all the figures presented in this paper, the radial coordinate $r$ has been rescaled using the mass of the boson star.

\begin{table*}[]
	\setlength{\tabcolsep}{10pt}
	\renewcommand{\arraystretch}{1}
	\centering
	\begin{tabular}{c c c c c c c}
		\hline\hline
		$\omega/\mu$ & $\phi_c$ &$A_c$ & $\mu M$ & $\mu R$  & $\mathcal{C}$ \\
		\hline\hline
		0.708457 & 0.0824 & 0.576784 & 0.269646 & 4.35805 & 0.061873 \\  
		0.550080 & 0.0880 & 0.416418 & 0.463052 & 4.35323 & 0.106370\\
		0.377812 & 0.0896 & 0.189622 & 0.957123 & 4.90497 & 0.195134\\
		0.320400 & 0.0952 & 0.069471 & 1.180510 & 4.74449 & 0.248817\\
		\hline\hline
	\end{tabular}
	\caption{Four different solitonic boson star configurations used in this paper. The parameter $\omega$ is the scalar field frequency, $\phi_c$ is the value of scalar field at the origin, $A_c$ is the value of the metric field at the origin, $\mu$ is the mass of the scalar field, $R$ is the radius of the star, $M$ is the total mass of the solitonic boson star, and $\mathcal{C} \equiv M/R$ is the compactness.}
	\flushleft
	\label{tab:soliton}
\end{table*}

\subsection{Radial kink model}
In order to study the evolution of the kink in the background of the boson star, we consider a probe real scalar field $\phi_k$ with the action\footnote{Here, in order to distinguish the kink scalar field $\phi_k$ from the scalar field in the above Eq. \eqref{eq:Phi}, we have added a subscript $k$ to the kink scalar field. Please be aware that the subscript $k$ here is different from the indices of tensors.}
\begin{equation}\label{S}
	S_k = \int d^4 x \sqrt{-g} \left[ \frac{1}{2} \, g_{\mu \nu}
	{\nabla}^{\mu} \phi_k {\nabla}^{\nu} \phi_k + V(\phi_k) \right] ,
\end{equation}
where the metric $g_{\mu\nu}$ is numerically obtained by solving the above Eqs. \eqref{eq:EQofMo1}-\eqref{eq:EQofMo3}, and the potential $V(\phi_k)$ is a double-well potential
\ber V(\phi_k) = \frac{1}{4} (\phi_k^2 - 1)^2, \eer 
which has minimums at $\phi_k=\pm 1$. Therefore, after breaking the $Z_2$ symmetry in the action \eqref{S}, the scalar field $\phi_k$ will settle down to $\phi_k=+1$ or $\phi_k=-1$, which in turn will form the solutions of kinks \cite{kibble1976topology,kibble1980some,zurek1985cosmological,ma2024asymmetric}. The equation of motion for the scalar field $\phi_k$ is 
\begin{equation}\label{eq:phimove}
	-\frac{1}{\sqrt{-g}} \,\partial_{\mu} \left( \sqrt{-g} \,g^{\mu \nu}
	\partial_{\nu} \phi_k \right) + \frac{\delta V(\phi_k)}{\delta \phi_{k}} = 0 .
\end{equation}
Since we work in the probe limit, the metric of the boson star will not be affected by the scalar field $\phi_k$. \footnote{It should be noted that the total action of the system is $S_{\text{tot}} = \mathcal{S} + \lambda S_k$, where $\mathcal{S}$ is the action of the background (i.e., Eq.\eqref{eq:action}) while $S_k$ is the kink scalar's action (i.e., Eq.\eqref{S}). The parameter $\lambda$ is the coupling constant. We work in the probe limit, which means that the coupling constant $\lambda$ is very small, and the backreaction of the real scalar fields to the background can be neglected. }

In this work, we are interested in the spherical solutions of kinks surrounding the boson star, and aim to study the impact of the boson star on the collapsing of the spherical kinks. Specifically, in the background of the spherical geometry, the Eq. \eqref{eq:phimove} can be re-expressed as 
\begin{align}\label{eq:phimove2}
	&\partial_t^2 \phi_k - \frac{1}{r^2} \,  G(r) \, \partial_r \left( r^2 G(r) \, \partial_r \phi_k \right)  
	- \frac{1}{r^2 \sin \theta} \,  A(r) \, \partial_{\theta} \Big( \sin \theta \, \partial_{\theta} \phi_k \Big)\nonumber \\	&- \frac{1}{r^2 \sin^2 \theta} \,  A(r) \, {\partial^2_{\varphi}  \phi_k } 
	+  A(r) \phi_k^3 -  A(r) \phi_k = 0 ,
\end{align}
where $G(r)=\sqrt{A(r)B(r)}$. Due to the spherical symmetry of the system, the polar and azimuthal angular derivative terms in the equation do not play a role.

\section{Radial kinks in the spherical boson stars}\label{sect4}

In this section we will study the evolution of kinks in the background of spherical boson stars, in particular in the background of massive and solitonic boson stars, respectively. The dynamics of kinks are governed by the Eq. \eqref{eq:phimove2}, with initial configurations of the kink given by 
\begin{equation}
	\label{eq:phiini1}
	\phi_k(0,r)= \tanh \left[ \frac{r-r_k(0)}{\sqrt{2 (1-v^2)}} \right] ,
\end{equation}
where $r_k(0)$ is the initial position of the kink at $t=0$ while $v$ represents the initial velocity of the kink. Throughout the paper, we always set $v=0$. This means that within the framework of the field model described by Eq. \eqref{eq:phimove2}, we assume initial conditions of a spherical radial kink with a given radius and zero initial velocity. The position of the kink is defined at the transition that $\phi_k$ changes its sign, i.e., $\phi_k(t,r_k)=0$. 

In numerics, we set the radial direction $r \in (0,100)$ and we discretize it into $4000$ grid points. Close to $r=0$, we impose the Neumann boundary conditions  $\partial_r \phi_k(t,0)=0$ to $\phi_k$, while at $r=100$ we fix $\phi_k(t,100)=1$. This requirement is consistent with the initial condition. The metric of the boson star background can be obtained by numerically solving Eqs. \eqref{eq:EQofMo1}-\eqref{eq:EQofMo3}. Configurations of the massive boson star and the solitonic boson star are shown in Fig. \ref{fig:massive} and Fig. \ref{fig:soliton}, respectively. Besides, we employ the fourth-order Runge-Kutta method in the time direction with a time step of $\Delta t=0.02$. In the radial direction $r$, we used the sixth-order difference method with $\Delta r=0.025$. To check the numerical stability, we also tested other time and spatial step sizes, such as $\Delta t=0.04$, $\Delta r=0.05$ or $\Delta t=0.01$, $\Delta r=0.0125$, and obtained consistent results. 

\begin{figure}[h]
	\centering
	\includegraphics[trim=3.3cm 8.3cm 2cm 9.5cm, clip=true, scale=0.49]{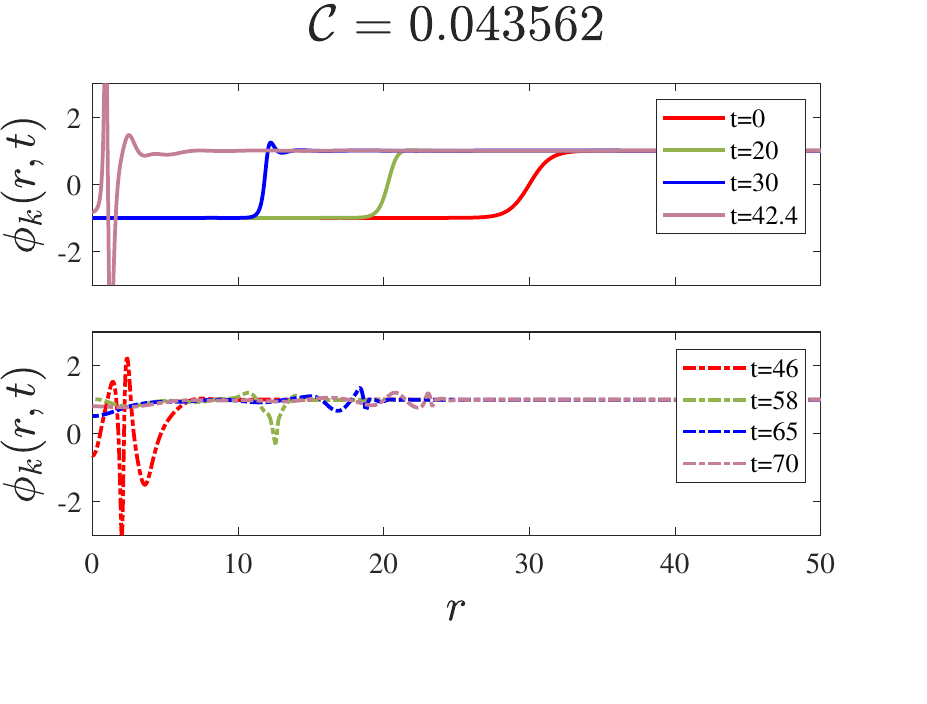}
	\put(-225,157){(a)}~
	\includegraphics[trim=3.3cm 8.3cm 2cm 9.5cm, clip=true, scale=0.49]{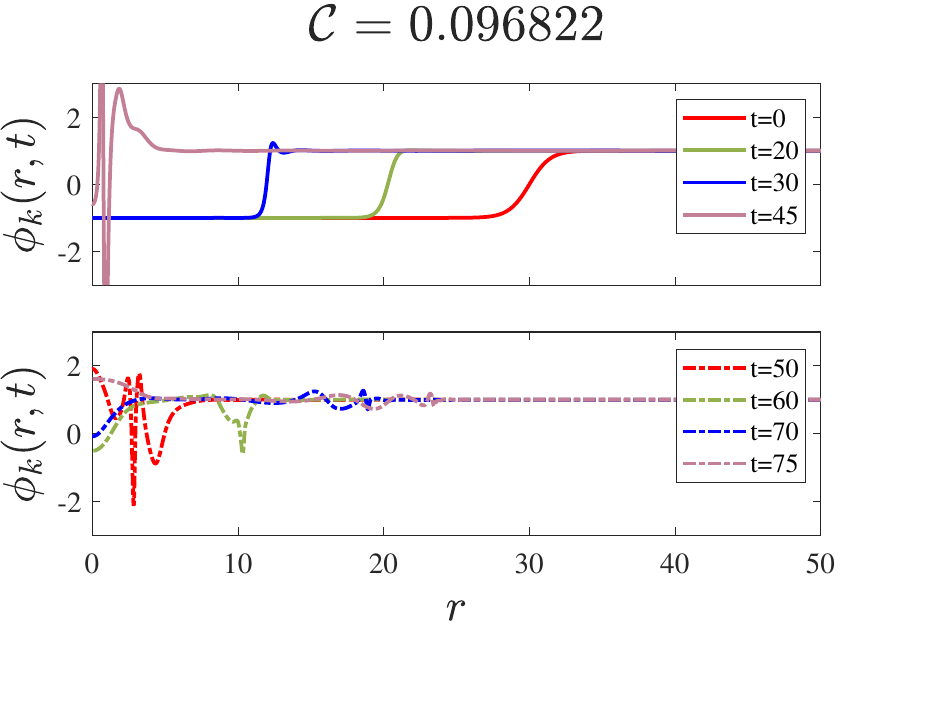}
	\put(-225,157){(b)}~\\
	\includegraphics[trim=3.3cm 9.5cm 2cm 9.5cm, clip=true, scale=0.49]{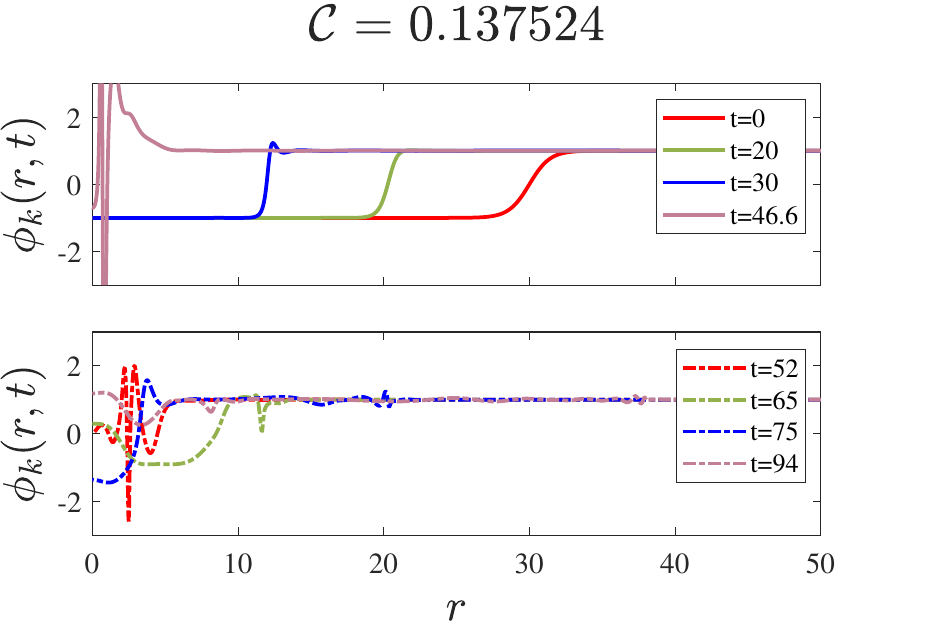}
	\put(-225,140){(c)}
	\includegraphics[trim=3.3cm 9.5cm 2cm 9.5cm, clip=true, scale=0.49]{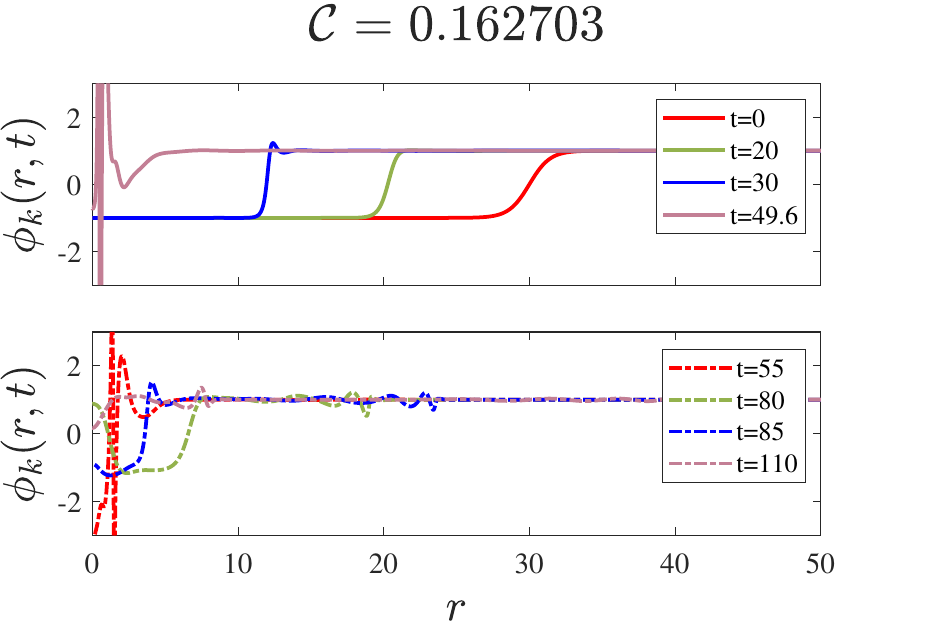}
	\put(-225,140){(d)}~
	\caption{Time evolutions of the scalar field $\phi_k$ in massive boson stars with four different compactness. In each panel, the scalar field shares the same initial configuration of the kink at position $r_k(0)=30$. In each panel, the upper plots show the kink evolutions (solid lines) before they collide with the origin, while the lower plots show the dynamics after the collision (dash-dotted lines). 
	}\label{fig:massivekink}
\end{figure}

\begin{figure}[htb]
	\centering
	\includegraphics[width=.4\textwidth]{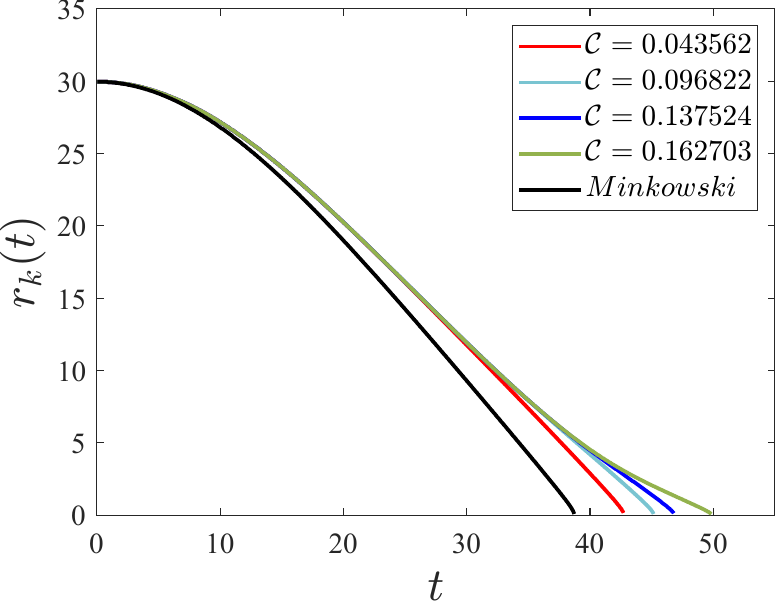}
	\caption{Time evolution of kink positions $r_k$ in massive boson star before they collide with the origin, with four different compactnesses. The black curve shows the reference evolution in Minkowskian spacetime. All cases share the same initial kink positions at $r_k(0)=30$ and initial zero velocities.}
	\label{fig:massivekinkP}
\end{figure}

\subsection{Kinks in massive boson stars}

We show the evolution of the kink in the background of massive boson stars in Fig. \ref{fig:massivekink} . In this figure, we have chosen four different massive boson stars with the related compactnesses as $\mathcal{C}=0.043562, 0.096822, 0.137524$, and $0.162703$, correspondingly shown in panels (a), (b), (c), and (d), respectively. In these panels, we choose the same initial position of the kink at $r_k(0)=30$ and the initial zero velocities. The upper panels exhibit the kinks' evolutions before they collide with the origin (solid lines), while the lower panels show the evolutions of $\phi_k(r)$ after the collision (dash-dotted lines). It is seen that the kinks reach the origin at approximately $t=42.4, 45, 46.6$, and $49.6$, respectively. This indicates that the compactness of the massive boson star will have impact to the collision rate of the kink: the larger the compactness is, the slower the kink collides to the origin. From Fig. \ref{fig:massive} we can observe that as the compactness is larger, the metric functions $A(r), B(r)$ will have deeper minimums and the scalar field $\phi(r)$ will have higher peak. We propose that these deeper and higher values in these metric and scalar functions may hinder the kinks to reach the origin. Thus, the larger the compactness is, the time for the kinks to collide with the origin will be. 

In order to study the precise relations between velocities of the kinks, we plot the kinks' positions $r_k(t)$ before colliding with the origin against time in Fig. \ref{fig:massivekinkP}.  In this figure, we compare the kinks' positions in massive boson stars with various compactness $\mathcal{C}$, as well as in the Minkowskian spacetime.\footnote{The numerical dynamics of the kinks in the Minkowskian spacetime can be found in Appdendix \ref{sect3}.} Specifically, we find that the velocity of the kinks will accelerate from zero at initial time because of our initial conditions $v=0$. For Minkowskian spacetime, after some time $t\approx 20$, the trajectory of the kink (black line) approaches a linear relation with time, indicating that the kink will reach the origin with a constant velocity. This phenomenon is easy to be understood since in Minkowskian spacetime there is nothing there to prevent the kink moving towards the origin. Therefore, the kink moves freely with constant velocity. However, for the massive boson stars, situations are complicated. If the compactness is small, such as $\mathcal{C}=0.043562$ (red line), the kink position will delay to arrive at the origin compared to the Minkowskian spacetime. This is because in the massive boson star background, the scalar field $\phi(r)$ is non-zero and the metric functions $A(r)$ and $B(r)$ are not constant (see Fig. \ref{fig:massive}), which may hinder the kink to move into the origin as we have discussed above. On the other hand, if the compactness is large, such as $\mathcal{C}=0.162703$ (green line), it is interesting to see that the kink velocity will decelerate in the late time around $t\approx 42$ (the green line bends at around $t\approx 42$ and then to have a smaller gradient). As we have explained above, in this case the metric functions and the scalar fields will have deeper minimum or higher peak, which may have stronger potentials to prevent the kink to move towards the origin. In summary, we can conclude that the presence of the massive boson star will slow down the kinks' velocity to reach the origin, and the compacter the boson star is, the stronger the effect will be. 

Moreover, in the lower plots of panels (a) and (b) of Fig. \ref{fig:massivekink}, we observe that after the kinks collide with the origin, the scalar fields will fluctuate dramatically and then quickly become ripples and dissipate into the background. However, from the lower plots of panels (c) and (d) of Fig.\ref{fig:massivekink}, we see that after the collision there still exists the possibility to form a new kink during the subsequent evolutions (cf. the green dash-dotted lines), although finally they will also dissipate into the background. This phenomenon is interesting which does not appear in the panels (a) and (b) of Fig. \ref{fig:massivekink}. Similar phenomena are also found in the Minkowskian spacetime in Appendix \ref{sect3}. The physical reasons are still unclear. However, from the numerical dynamics we can draw a conclusion that this new formed kink is highly related to the compactness of the massive boson stars, i.e., when a kink collides with a boson star with very high compactness, it is possible to form a new kink after the collision.

\subsection{Kinks in solitonic boson stars}

\begin{figure}[h]
	\centering
	\includegraphics[trim=3.3cm 8.3cm 2cm 9.5cm, clip=true, scale=0.49]{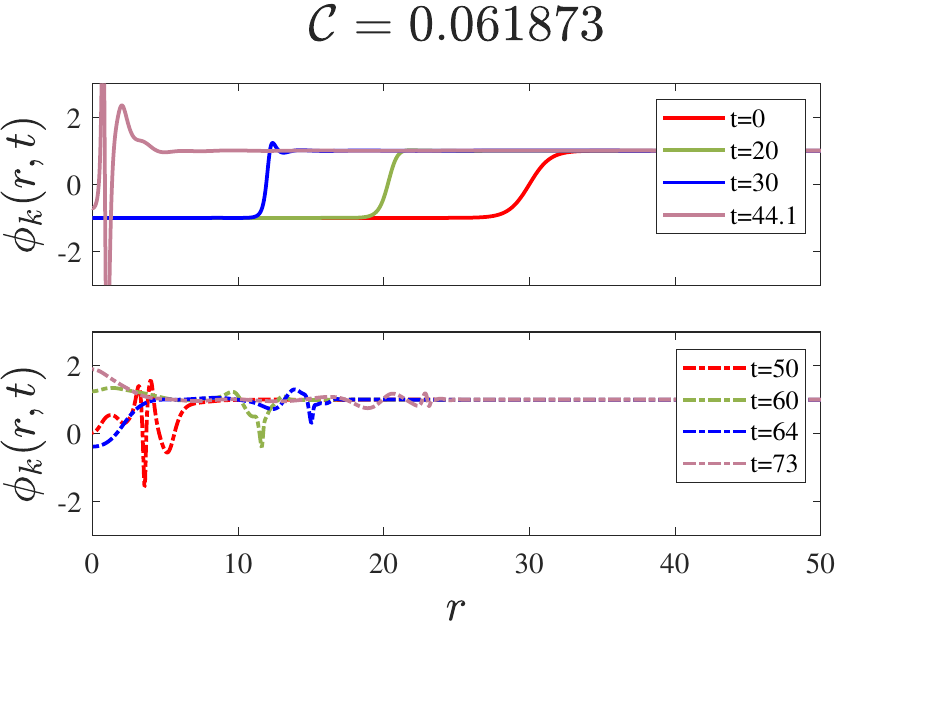}
	\put(-225,157){(a)}~
	\includegraphics[trim=3.3cm 8.3cm 2cm 9.5cm, clip=true, scale=0.49]{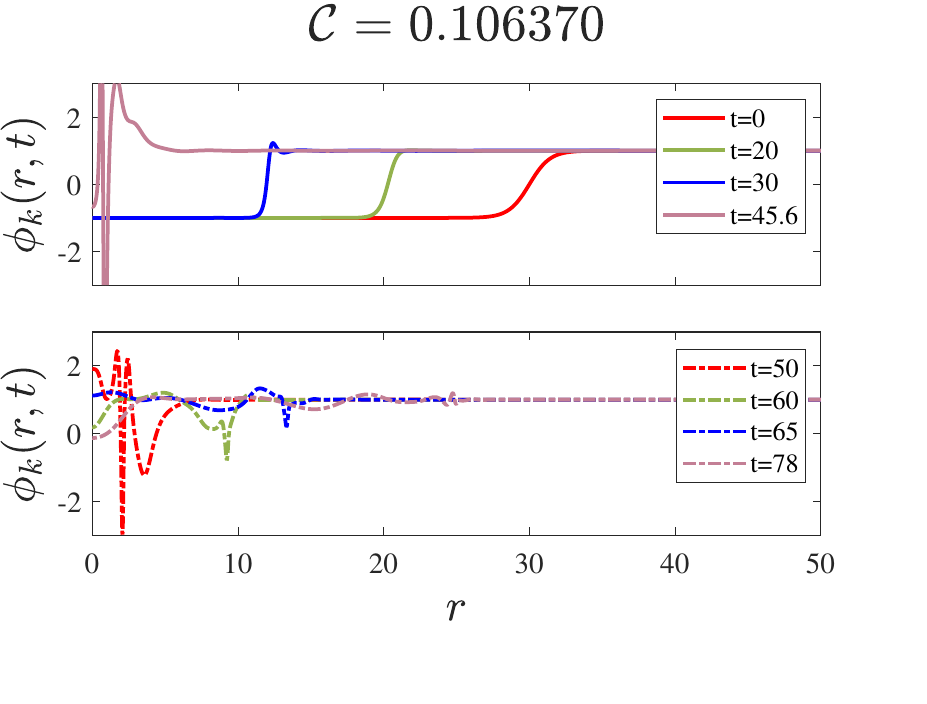}
	\put(-225,157){(b)}~\\
	\includegraphics[trim=3.3cm 9.5cm 2cm 9.5cm, clip=true, scale=0.49]{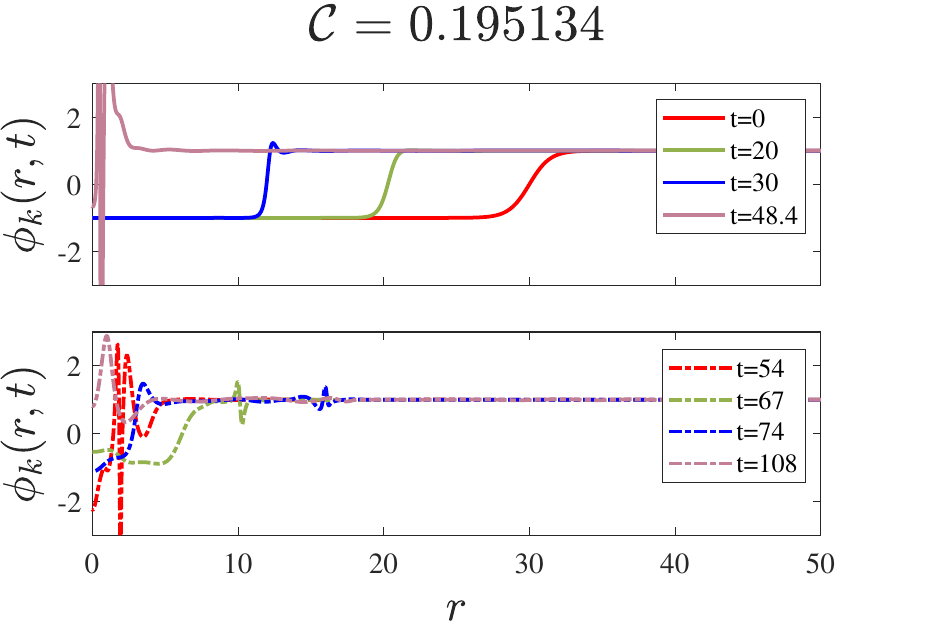}
	\put(-225,140){(c)}
	\includegraphics[trim=3.3cm 9.5cm 2cm 9.5cm, clip=true, scale=0.49]{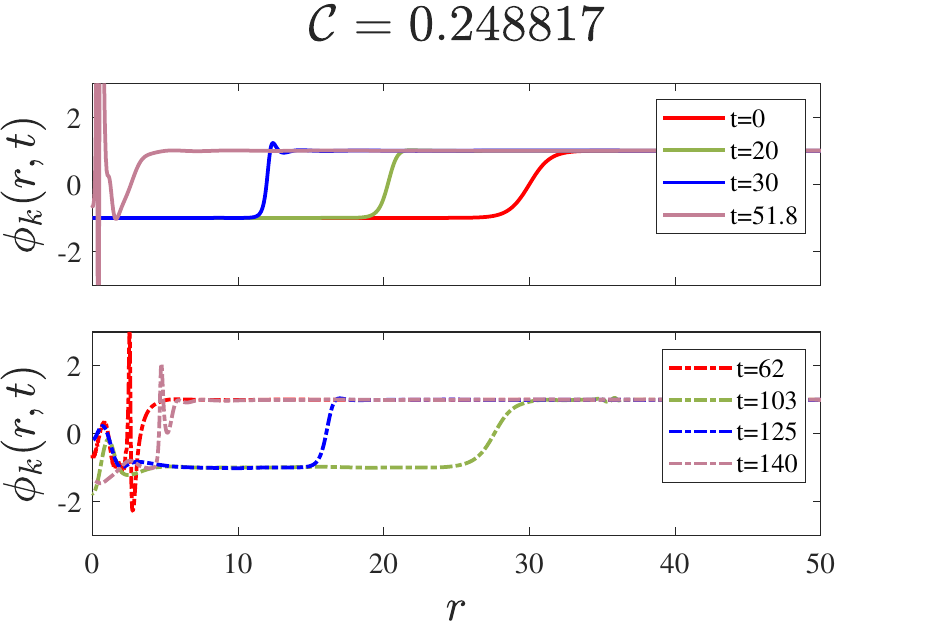}
	\put(-225,140){(d)}~
	\caption{Time evolutions of the scalar field $\phi_k$ in solitonic boson stars with four different compactness. In each panel, the scalar field shares the same initial configuration of the kink at position $r_k(0)=30$. In each panel, the upper plots show the kink evolutions (solid lines) before they collide with the origin, while the lower plots show the dynamics after the collision (dash-dotted lines). 
	}\label{fig:solitonkink}
\end{figure}

\begin{figure}[h]
	\centering
	\includegraphics[width=0.4\textwidth]{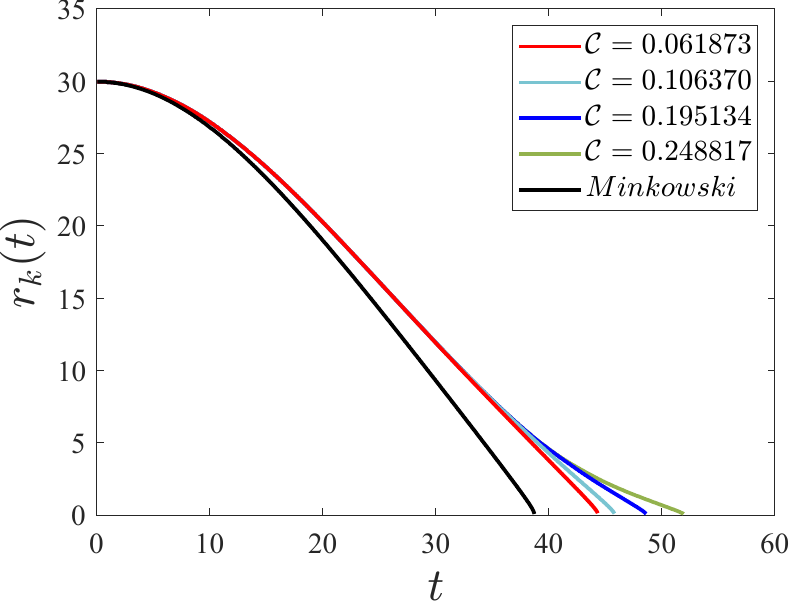}
	\caption{Time evolution of kink positions $r_k$ in solitonic boson star before they collide with the origin, with four different compactnesses. The black curve shows the reference evolution in Minkowskian spacetime. All cases share the same initial kink positions at $r_k(0)=30$ and initial zero velocities.}
	\label{fig:solitonkinkP}
\end{figure}

In this subsection we will further study the kink evolutions in the spherically symmetric solitonic boson stars, and compare it with those in the Minkowskian spacetime. In particular, we investigate the kink evolutions in the solitonic boson stars with four different compactness, i.e., $\mathcal{C}=0.061873, 0.106370, 0.195134$ and $0.248817$, see Fig. \ref{fig:solitonkink}. In the four panels of Fig. \ref{fig:solitonkink}, we set the initial positions of the kinks all at $r_k(0)=30$ and the zero velocities. The upper plots of each panel show the kink evolution before they collide with the origin, while the lower plots show the dynamics after the collision. We see that the dynamics of the kinks are very similar to those in the massive boson stars above:
\begin{itemize}
\item The larger the compactness is, the longer for the kinks to reach the origin. It can be obviously seen from the solid purple lines that the colliding time for the kinks are $t=44.1, 45.6, 48.4$ and $51.8$ respectively from the panels (a)-(d) in Fig. \ref{fig:solitonkink}. In order to study the velocities of the kinks to reach the origin in detail, we plot the time evolutions of the kink positions $r_k(t)$ in Fig. \ref{fig:solitonkinkP}. From this figure, we can see that compared to the Minkowskian spacetime (black line), the kinks are slower to reach the origin in the solitonic boson stars. Besides, the larger the compactness is, the slower the kinks are especially at late time. The physical reason resembles those in the massive boson stars. In Fig. \ref{fig:soliton} we see that as the compactness is larger, the metric function $A(r)$ and $B(r)$ will have deeper minimums, while the scalar field $\phi(r)$ will have higher peak. These minimums and peaks will hinder the kinks to collide with the origin. 
\item As the compactness becomes larger, there is possibility to form a new kink after the collision. As the compactness is small, see for instance the lower plots in panels (a) and (b) in Fig. \ref{fig:solitonkink}, after the collision the scalar field will fluctuate and become ripples and then dissipate into the background. However, as the compactness is large, see the lower plots in panels (c) and (d) in Fig. \ref{fig:solitonkink}, there exist new kinks (green dash-dotted lines) after the collision. Moreover, in panel (d) which has the largest compactness, we can even find that after the collision the kink can move outward far away from the origin, and then move towards the origin again in the late time, although finally it will become ripples and dissipate into the background. 
\end{itemize}

It is worth noting that in the Minkowskian spacetime, there is also possibility to form a new kink in the subsequent dynamics after the collision. However, there the mechanisms are different, since Minkowskian spacetime does not have the compactness like the boson stars. In Appendix \ref{sect3},  we will find that if the initial position of the kink is further away from the origin, it has more chance to form a new kink after the collision.

\section{Conclusions and Discussions}\label{sect5}
In this work, we studied the evolutions of the radial kinks in the background of spherical massive boson stars and solitonic boson stars. The results showed that the existence of boson stars would delay the collision of the kinks with the origin, comparing to the Minkowskian spacetime. Besides, in both massive and solitonic boson stars, the larger the compactness is, the longer time the kinks would take to collide with the origin. After the collision, the scalar fields would fluctuate and become ripples and then dissipate into the background. However, for both boson stars with larger compactness, we could see the transient new formed kinks after the collision, although they would finally dissipated into the background as well.

In the solitonic boson star with very high compactness, we even saw a new formed kink which moved much far away from the origin after the collision. Based on this, we may conjecture that the process of the kink entering the interior of the boson star and subsequently moving outward might carry some information from within the boson star. Therefore, it could potentially provide some clues for the internal structures of the dense celestial objects. On the other hand, the kinks have comparable sizes to the boson stars and decay rapidly in the late time. Therefore, extracting information from the ripples generated by the dissipation of the kink is also a promising option. This is a very strong motivation for our work. Moreover, if we work in the background of black holes, this phenomenon may also extract some information from the interior of black holes, which may provide another avenue on studying the black hole information. 

Of course, our work still has some limitations. One limitation is that the physical reasoning for the impact of the compactness on the new formed kinks after collision is still vague. We could only observe it from the numerical results at the present time, rather than from the analytical way since the dynamics is very complicated. Another limitations is that currently for simplicity we are working in the probe limit, rather than considering the backreacton effect of the kinks to the background of the boson stars. In order to be more precise, working with the backreaction remains a subject for future investigation.

\section*{Acknowledgements}
This work was partially supported by the National Natural Science Foundation of China (Grants No.12175008).

\appendix
\setcounter{figure}{0}
\renewcommand{\thefigure}{A\arabic{figure}}

\section{Kink evolution in Minkowskian spacetime}\label{sect3}

In the appendix, we will show the evolution of the spherical kinks in the Minkowskian spacetime. The initial configuration of $\phi_k$ is shown in Eq.\eqref{eq:phiini1} in the main text and the initial velocity of the kink is $v=0$. In numerics, we set the radial direction $r \in (0,100)$ and we discretize it into $4000$ grid points. Close to $r=0$, we impose the Neumann boundary conditions to $\phi_k$, while at $r=100$ we fix $\phi_k(t,100)=1$. This requirement is consistent with the initial condition.

\begin{figure}[h]
	\centering
	\includegraphics[trim=3.3cm 8.3cm 2cm 9.5cm, clip=true, scale=0.49]{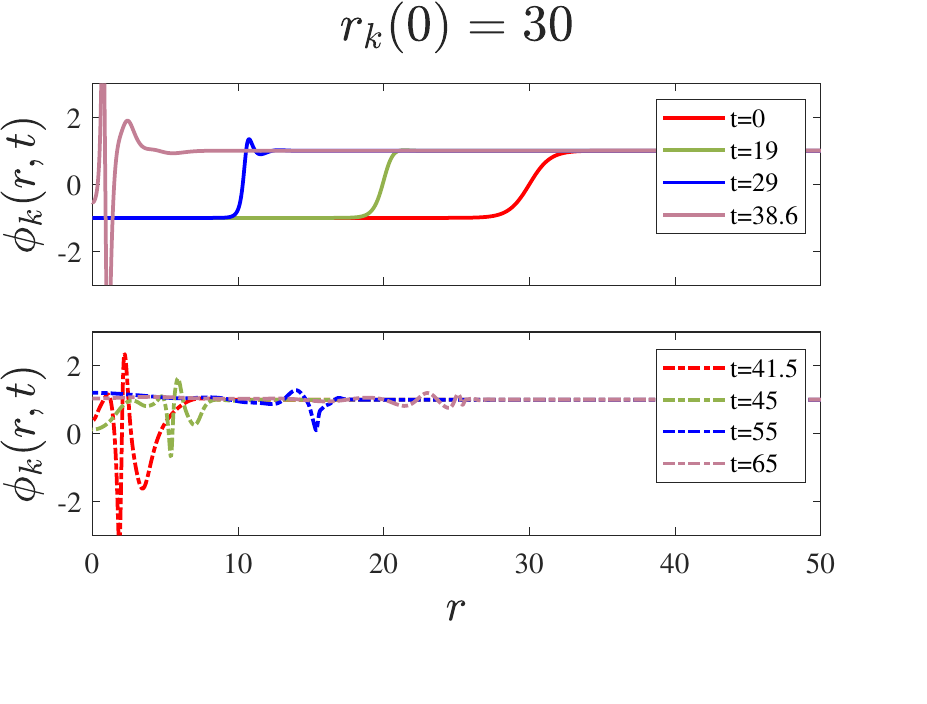}
	\put(-225,157){(a)}~
	\includegraphics[trim=3.3cm 8.3cm 2cm 9.5cm, clip=true, scale=0.49]{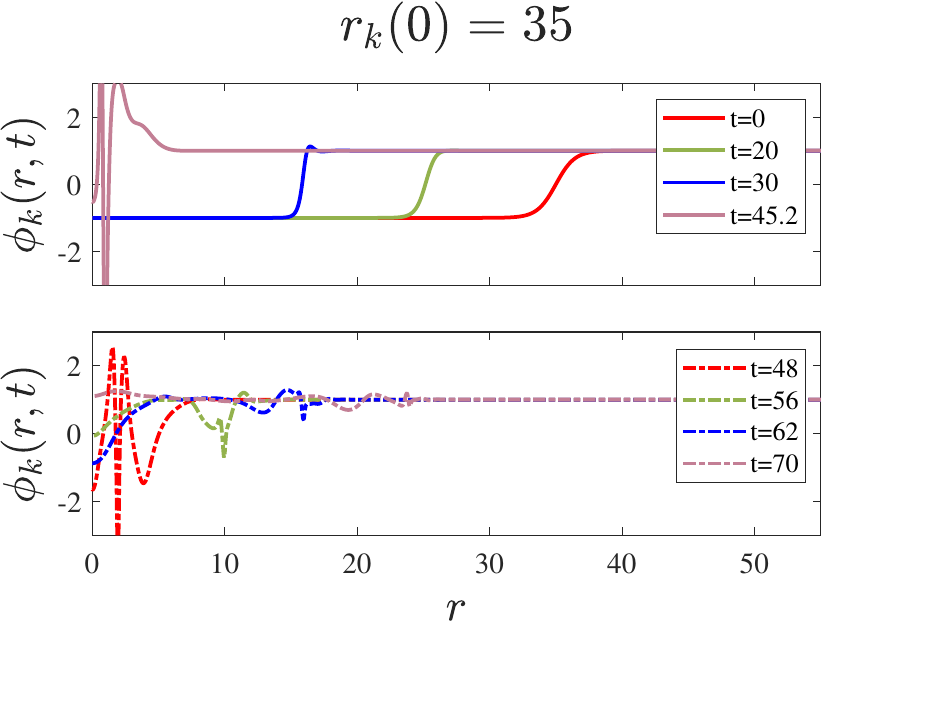}
	\put(-225,157){(b)}~\\
	\includegraphics[trim=3.3cm 9.5cm 2cm 9.5cm, clip=true, scale=0.49]{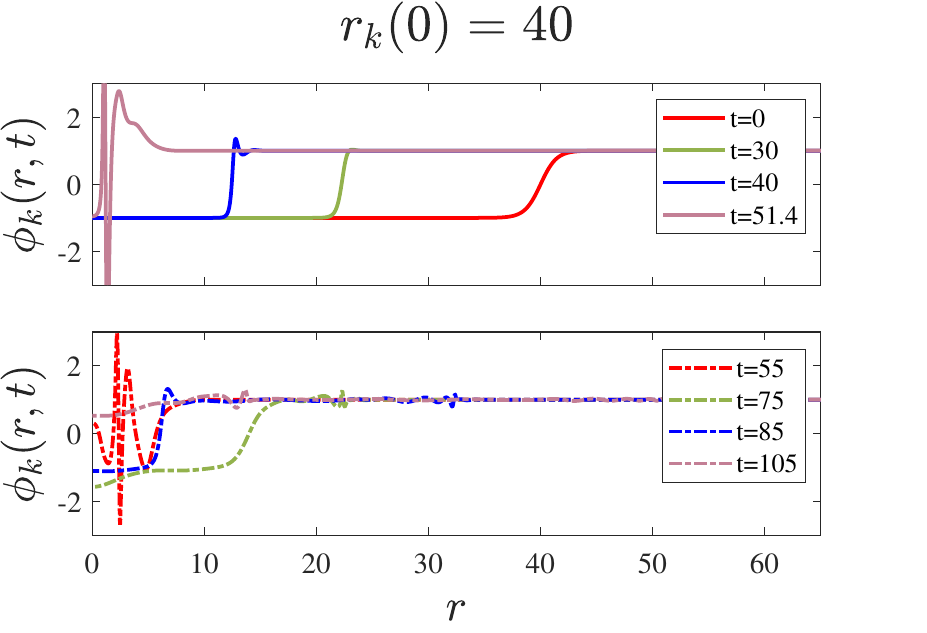}
	\put(-225,140){(c)}
	\includegraphics[trim=3.3cm 9.5cm 2cm 9.5cm, clip=true, scale=0.49]{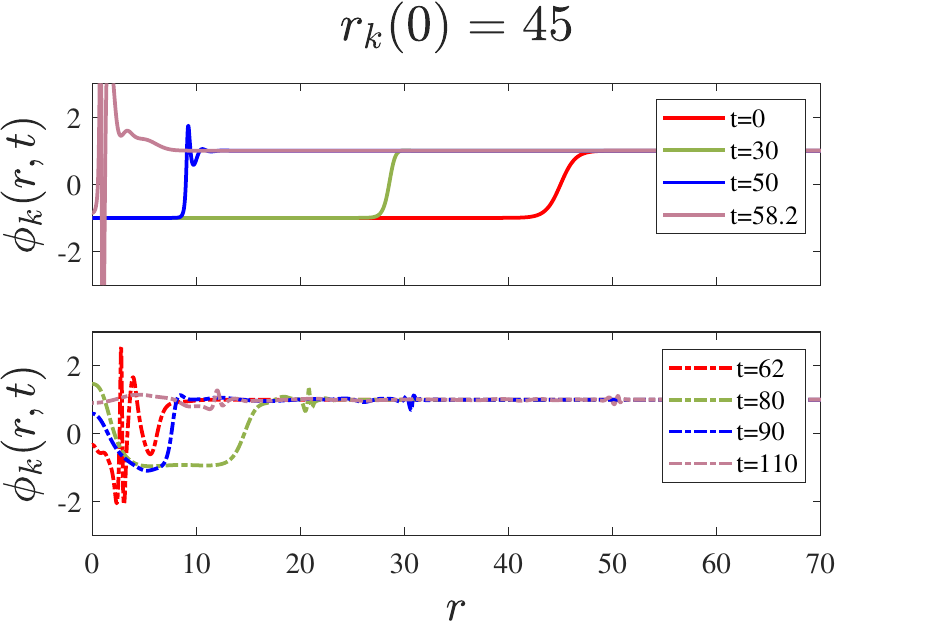}
	\put(-225,140){(d)}~
	\caption{Time evolution of the scalar field $\phi_k$ in (3+1)-dimensional Minkowski spacetime. The four panels illustrate different initial kink positions: (a) $r_k(0)$=30, (b) $r_k(0)$=35, (c) $r_k(0)$=40, and (d) $r_k(0)$=45. 
	}\label{fig:MinkowKink}
\end{figure}

In Fig. \ref{fig:MinkowKink}, we present the evolution of the scalar field $\phi_k$ in the Minkowskian spacetime with various initial positions $r_k(0)$. The initial configurations of the scalar field are set to be kink profiles as in Eq.\eqref{eq:phiini1}. In panels (a), (b), (c) and (d), the initial kink positions are $r_k(0)=30$, $r_k(0)=35$, $r_k(0)=40$, and $r_k(0)=45$, respectively. In each panel, two subplots are provided: the upper subplot displays the scalar field configuration of the kink before it collides with the origin, while the lower subplot shows the scalar field configuration after colliding. It is straightforward to see that the further kink position is, the longer it takes for the kink to reach the origin. As shown in the figure, for initial kink positions of $r_k(0)=30$, $r_k(0)=35$, $r_k(0)=40$, and $r_k(0)=45$, the times for the kink to reach the  origin are approximately $t=38.6$, $t=45.2$, $t=51.4$ and $t=58.2$, respectively. 

Furthermore, we observe that when the initial position of the kink is relatively close to the origin, as shown in the Fig. \ref{fig:MinkowKink} (a) and Fig. \ref{fig:MinkowKink} (b), after the kink collides with the origin, it rapidly fluctuates and subsequently dissipates into ripples in the background. However, when the initial position of the kink is away from the origin, a distinct behavior emerges. As seen in the field configurations at $t=75$ in Fig. \ref{fig:MinkowKink} (c) and $t=80$ in Fig. \ref{fig:MinkowKink} (d), after the kink collides with the origin, it does not immediately dissipate into the background, but instead it has the opportunity to form a new kink (green dash-dotted line). Although the newly formed kink will eventually  collides with the origin again and dissipates into ripples in the background, this intermediate process of kink formation is evident and comparable with those in the Fig. \ref{fig:MinkowKink} (a) and Fig. \ref{fig:MinkowKink} (b).

\normalem
\bibliographystyle{ieeetr}
\bibliography{ref1.bib}

\begin{thebibliography}{10}

\bibitem{kibble1982phase}
T.~W.~B. Kibble and G.~In, ``Phase transitions in the early universe,'' {\em
  Quantum Structure of Space and Time}, p.~391, 1982.

\bibitem{vilenkin1994cosmic}
A.~Vilenkin, A.~Vilenkin, and E.~Shellard, {\em Cosmic strings and other
  topological defects}.
\newblock Cambridge University Press, 1994.

\bibitem{vachaspati2006kinks}
T.~Vachaspati, {\em Kinks and domain walls: An introduction to classical and
  quantum solitons}.
\newblock Cambridge University Press, 2006.

\bibitem{pismen1999vortices}
L.~M. Pismen, {\em Vortices in nonlinear fields: From liquid crystals to
  superfluids, from non-equilibrium patterns to cosmic strings}, vol.~100.
\newblock Oxford University Press, 1999.

\bibitem{manton2004topological}
N.~Manton and P.~Sutcliffe, {\em Topological solitons}.
\newblock Cambridge University Press, 2004.

\bibitem{bunkov2000topological}
Y.~M. Bunkov and H.~Godfrin, {\em Topological defects and the non-equilibrium
  dynamics of symmetry breaking phase transitions}, vol.~549.
\newblock Springer Science \& Business Media, 2000.

\bibitem{kibble1976topology}
T.~W. Kibble, ``Topology of cosmic domains and strings,'' {\em Journal of
  Physics A: Mathematical and General}, vol.~9, no.~8, p.~1387, 1976.

\bibitem{vilenkin1982cosmic}
A.~Vilenkin and A.~E. Everett, ``Cosmic strings and domain walls in models with
  goldstone and pseudo-goldstone bosons,'' {\em Physical Review Letters},
  vol.~48, no.~26, p.~1867, 1982.

\bibitem{vilenkin1985cosmic}
A.~Vilenkin, ``Cosmic strings and domain walls,'' {\em Physics reports},
  vol.~121, no.~5, pp.~263--315, 1985.

\bibitem{press1989dynamical}
W.~H. Press, B.~S. Ryden, and D.~N. Spergel, ``Dynamical evolution of domain
  walls in an expanding universe,'' {\em Astrophysical Journal, Part 1 (ISSN
  0004-637X), vol. 347, Dec. 15, 1989, p. 590-604. Research supported by NASA
  and Alfred P. Sloan Foundation.}, vol.~347, pp.~590--604, 1989.

\bibitem{avelino2008dynamics}
P.~Avelino, C.~Martins, J.~Menezes, R.~Menezes, and J.~Oliveira, ``Dynamics of
  domain wall networks with junctions,'' {\em Physical Review D—Particles,
  Fields, Gravitation, and Cosmology}, vol.~78, no.~10, p.~103508, 2008.

\bibitem{friedland2003domain}
A.~Friedland, H.~Murayama, and M.~Perelstein, ``Domain walls as dark energy,''
  {\em Physical Review D}, vol.~67, no.~4, p.~043519, 2003.

\bibitem{derevianko2014hunting}
A.~Derevianko and M.~Pospelov, ``Hunting for topological dark matter with
  atomic clocks,'' {\em Nature Physics}, vol.~10, no.~12, pp.~933--936, 2014.

\bibitem{roberts2017search}
B.~M. Roberts, G.~Blewitt, C.~Dailey, M.~Murphy, M.~Pospelov, A.~Rollings,
  J.~Sherman, W.~Williams, and A.~Derevianko, ``Search for domain wall dark
  matter with atomic clocks on board global positioning system satellites,''
  {\em Nature communications}, vol.~8, no.~1, p.~1195, 2017.

\bibitem{kalaydzhyan2017extracting}
T.~Kalaydzhyan and N.~Yu, ``Extracting dark matter signatures from atomic clock
  stability measurements,'' {\em Physical Review D}, vol.~96, no.~7, p.~075007,
  2017.

\bibitem{stadnik2014searching}
Y.~Stadnik and V.~Flambaum, ``Searching for topological defect dark matter via
  nongravitational signatures,'' {\em Physical review letters}, vol.~113,
  no.~15, p.~151301, 2014.

\bibitem{pustelny2013global}
S.~Pustelny, D.~F. Jackson~Kimball, C.~Pankow, M.~P. Ledbetter, P.~Wlodarczyk,
  P.~Wcislo, M.~Pospelov, J.~R. Smith, J.~Read, W.~Gawlik, {\em et~al.}, ``The
  global network of optical magnetometers for exotic physics (gnome): A novel
  scheme to search for physics beyond the standard model,'' {\em Annalen der
  Physik}, vol.~525, no.~8-9, pp.~659--670, 2013.

\bibitem{afach2021search}
S.~Afach, B.~C. Buchler, D.~Budker, C.~Dailey, A.~Derevianko, V.~Dumont, N.~L.
  Figueroa, I.~Gerhardt, Z.~D. Gruji{\'c}, H.~Guo, {\em et~al.}, ``Search for
  topological defect dark matter with a global network of optical
  magnetometers,'' {\em Nature Physics}, vol.~17, no.~12, pp.~1396--1401, 2021.

\bibitem{mcnally2020constraining}
R.~L. McNally and T.~Zelevinsky, ``Constraining domain wall dark matter with a
  network of superconducting gravimeters and ligo,'' {\em The European Physical
  Journal D}, vol.~74, no.~4, p.~61, 2020.

\bibitem{hall2018laser}
E.~D. Hall, R.~X. Adhikari, V.~V. Frolov, H.~M{\"u}ller, and M.~Pospelov,
  ``Laser interferometers as dark matter detectors,'' {\em Physical Review D},
  vol.~98, no.~8, p.~083019, 2018.

\bibitem{grote2019novel}
H.~Grote and Y.~Stadnik, ``Novel signatures of dark matter in
  laser-interferometric gravitational-wave detectors,'' {\em Physical Review
  Research}, vol.~1, no.~3, p.~033187, 2019.

\bibitem{jaeckel2021probing}
J.~Jaeckel, S.~Schenk, and M.~Spannowsky, ``Probing dark matter clumps, strings
  and domain walls with gravitational wave detectors,'' {\em The European
  Physical Journal C}, vol.~81, no.~9, p.~828, 2021.

\bibitem{kevrekidis2018planar}
P.~G. Kevrekidis, I.~Danaila, J.-G. Caputo, and R.~Carretero-Gonz{\'a}lez,
  ``Planar and radial kinks in nonlinear klein-gordon models: Existence,
  stability, and dynamics,'' {\em Physical Review E}, vol.~98, no.~5,
  p.~052217, 2018.

\bibitem{carretero2022kink}
R.~Carretero-Gonz{\'a}lez, L.~Cisneros-Ake, R.~Decker, G.~Koutsokostas, D.~J.
  Frantzeskakis, P.~Kevrekidis, and D.~J. Ratliff, ``Kink--antikink stripe
  interactions in the two-dimensional sine--gordon equation,'' {\em
  Communications in Nonlinear Science and Numerical Simulation}, vol.~109,
  p.~106123, 2022.

\bibitem{caputo2013radial}
J.-G. Caputo and M.~P. S{\o}rensen, ``Radial sine-gordon kinks as sources of
  fast breathers,'' {\em Physical Review E—Statistical, Nonlinear, and Soft
  Matter Physics}, vol.~88, no.~2, p.~022915, 2013.

\bibitem{maldacena1999large}
J.~Maldacena, ``The large-n limit of superconformal field theories and
  supergravity,'' {\em International journal of theoretical physics}, vol.~38,
  no.~4, pp.~1113--1133, 1999.

\bibitem{witten1998anti}
E.~Witten, ``Anti de sitter space and holography,'' {\em arXiv preprint
  hep-th/9802150}, 1998.

\bibitem{li2023black}
Z.-H. Li, H.-Q. Shi, and H.-Q. Zhang, ``From black hole to one-dimensional
  chain: Parity symmetry breaking and kink formation,'' {\em Physical Review
  D}, vol.~108, no.~10, p.~106015, 2023.

\bibitem{ma2025universal}
T.-C. Ma, H.-Q. Shi, H.-Q. Zhang, and A.~del Campo, ``Universal critical
  holography and domain wall formation,'' {\em Physical Review Research},
  vol.~7, no.~1, p.~013096, 2025.

\bibitem{arodz1998expansion}
H.~Arodz, ``Expansion in the width and collective dynamics of a domain wall,''
  {\em Nuclear Physics B}, vol.~509, no.~1-2, pp.~273--293, 1998.

\bibitem{dobrowolski2008construction}
T.~Dobrowolski, ``Construction of curved domain walls,'' {\em Physical Review
  E—Statistical, Nonlinear, and Soft Matter Physics}, vol.~77, no.~5,
  p.~056608, 2008.

\bibitem{dobrowolski2009kink}
T.~Dobrowolski, ``Kink motion in a curved josephson junction,'' {\em Physical
  Review E—Statistical, Nonlinear, and Soft Matter Physics}, vol.~79, no.~4,
  p.~046601, 2009.

\bibitem{gatlik2021modeling}
J.~Gatlik and T.~Dobrowolski, ``Modeling kink dynamics in the sine--gordon
  model with position dependent dispersive term,'' {\em Physica D: Nonlinear
  Phenomena}, vol.~428, p.~133061, 2021.

\bibitem{gorria2004kink}
C.~Gorria, Y.~B. Gaididei, M.~P. S{\o}rensen, P.~L. Christiansen, and J.~G.
  Caputo, ``Kink propagation and trapping in a two-dimensional curved josephson
  junction,'' {\em Physical Review B}, vol.~69, no.~13, p.~134506, 2004.

\bibitem{shi2024topologically}
H.-Q. Shi, T.-C. Ma, and H.-Q. Zhang, ``Topologically protected metastable
  states in classical dynamics,'' {\em Chaos, Solitons \& Fractals}, vol.~182,
  p.~114789, 2024.

\bibitem{moderski2003thick}
R.~Moderski and M.~Rogatko, ``Thick domain walls and charged dilaton black
  holes,'' {\em Physical Review D}, vol.~67, no.~2, p.~024006, 2003.

\bibitem{morisawa2003thick}
Y.~Morisawa, D.~Ida, A.~Ishibashi, and K.-i. Nakao, ``Thick domain walls around
  a black hole,'' {\em Physical Review D}, vol.~67, no.~2, p.~025017, 2003.

\bibitem{moderski2004reissner}
R.~Moderski and M.~Rogatko, ``Reissner-nordstr{\"o}m black holes and thick
  domain walls,'' {\em Physical Review D}, vol.~69, no.~8, p.~084018, 2004.

\bibitem{moderski2006thick}
R.~Moderski and M.~Rogatko, ``Thick domain walls in ads black hole
  spacetimes,'' {\em Physical Review D—Particles, Fields, Gravitation, and
  Cosmology}, vol.~74, no.~4, p.~044002, 2006.

\bibitem{ficek2018planar}
F.~Ficek and P.~Mach, ``Planar domain walls in black hole spacetimes,'' {\em
  Physical Review D}, vol.~97, no.~4, p.~044012, 2018.

\bibitem{caputo2024radial}
J.-G. Caputo, T.~Dobrowolski, J.~Gatlik, and P.~G. Kevrekidis, ``Radial kinks
  in a schwarzschild-like geometry,'' {\em Physical Review D}, vol.~110,
  no.~12, p.~125025, 2024.

\bibitem{hawking1970singularities}
S.~W. Hawking and R.~Penrose, ``The singularities of gravitational collapse and
  cosmology,'' {\em Proceedings of the Royal Society of London. A. Mathematical
  and Physical Sciences}, vol.~314, no.~1519, pp.~529--548, 1970.

\bibitem{penrose1965gravitational}
R.~Penrose, ``Gravitational collapse and space-time singularities,'' {\em
  Physical Review Letters}, vol.~14, no.~3, p.~57, 1965.

\bibitem{senovilla1998singularity}
J.~M. Senovilla, ``Singularity theorems and their consequences,'' {\em General
  Relativity and Gravitation}, vol.~30, no.~5, pp.~701--848, 1998.

\bibitem{bardeen1968non}
J.~Bardeen, ``Non-singular general relativistic gravitational collapse,'' in
  {\em Proceedings of the 5th International Conference on Gravitation and the
  Theory of Relativity}, p.~87, 1968.

\bibitem{hayward2006formation}
S.~A. Hayward, ``Formation and evaporation of nonsingular black holes,'' {\em
  Physical review letters}, vol.~96, no.~3, p.~031103, 2006.

\bibitem{ayon1998regular}
E.~Ayon-Beato and A.~Garcia, ``Regular black hole in general relativity coupled
  to nonlinear electrodynamics,'' {\em Physical review letters}, vol.~80,
  no.~23, p.~5056, 1998.

\bibitem{berej2006regular}
W.~Berej, J.~Matyjasek, D.~Tryniecki, and M.~Woronowicz, ``Regular black holes
  in quadratic gravity,'' {\em General Relativity and Gravitation}, vol.~38,
  no.~5, pp.~885--906, 2006.

\bibitem{colpi1986boson}
M.~Colpi, S.~L. Shapiro, and I.~Wasserman, ``Boson stars: Gravitational
  equilibria of self-interacting scalar fields,'' {\em Physical review
  letters}, vol.~57, no.~20, p.~2485, 1986.

\bibitem{lee1987soliton}
T.~Lee, ``Soliton stars and the critical masses of black holes,'' {\em Physical
  Review D}, vol.~35, no.~12, p.~3637, 1987.

\bibitem{vincent2016imaging}
F.~Vincent, Z.~Meliani, P.~Grandclement, E.~Gourgoulhon, and O.~Straub,
  ``Imaging a boson star at the galactic center,'' {\em Classical and Quantum
  Gravity}, vol.~33, no.~10, p.~105015, 2016.

\bibitem{ma2023hybrid}
T.-X. Ma, C.~Liang, J.~Yang, and Y.-Q. Wang, ``Hybrid proca-boson stars,'' {\em
  Physical Review D}, vol.~108, no.~10, p.~104011, 2023.

\bibitem{ma2025boson}
T.-X. Ma, T.-F. Fang, and Y.-Q. Wang, ``Boson stars and their frozen states in
  an infinite tower of higher-derivative gravity,'' {\em The European Physical
  Journal C}, vol.~85, no.~5, p.~542, 2025.

\bibitem{brito2016proca}
R.~Brito, V.~Cardoso, C.~A. Herdeiro, and E.~Radu, ``Proca stars: Gravitating
  bose--einstein condensates of massive spin 1 particles,'' {\em Physics
  Letters B}, vol.~752, pp.~291--295, 2016.

\bibitem{landea2016charged}
I.~S. Landea and F.~Garc{\'\i}a, ``Charged proca stars,'' {\em Physical Review
  D}, vol.~94, no.~10, p.~104006, 2016.

\bibitem{rosa2023imaging}
J.~L. Rosa, C.~F. Macedo, and D.~Rubiera-Garcia, ``Imaging compact boson stars
  with hot spots and thin accretion disks,'' {\em Physical Review D}, vol.~108,
  no.~4, p.~044021, 2023.

\bibitem{johnson2020universal}
M.~D. Johnson, A.~Lupsasca, A.~Strominger, G.~N. Wong, S.~Hadar, D.~Kapec,
  R.~Narayan, A.~Chael, C.~F. Gammie, P.~Galison, {\em et~al.}, ``Universal
  interferometric signatures of a black hole’s photon ring,'' {\em Science
  advances}, vol.~6, no.~12, p.~eaaz1310, 2020.

\bibitem{olmo2023shadows}
G.~J. Olmo, J.~L. Rosa, D.~Rubiera-Garcia, and D.~S.-C. Gomez, ``Shadows and
  photon rings of regular black holes and geonic horizonless compact objects,''
  {\em Classical and Quantum Gravity}, vol.~40, no.~17, p.~174002, 2023.

\bibitem{pitz2023generating}
S.~L. Pitz and J.~Schaffner-Bielich, ``Generating ultracompact boson stars with
  modified scalar potentials,'' {\em Physical Review D}, vol.~108, no.~10,
  p.~103043, 2023.

\bibitem{he2025observation}
K.-J. He, G.-P. Li, C.-Y. Yang, and X.-X. Zeng, ``The observation image of a
  soliton boson star illuminated by various accretions,'' {\em arXiv preprint
  arXiv:2502.16623}, 2025.

\bibitem{zeng2025optical}
X.-X. Zeng, H.~Ye, K.-J. He, and H.~Yu, ``Optical images of massive boson stars
  with nonlinear electrodynamics,'' {\em arXiv preprint arXiv:2507.11583},
  2025.

\bibitem{zeng2025polarization}
X.-X. Zeng, C.-Y. Yang, H.~Yu, and K.-J. He, ``Polarization images of solitonic
  boson stars,'' {\em arXiv preprint arXiv:2508.11992}, 2025.

\bibitem{jetzer1992boson}
P.~Jetzer, ``Boson stars,'' {\em Physics Reports}, vol.~220, no.~4,
  pp.~163--227, 1992.

\bibitem{gleiser1988stability}
M.~Gleiser, ``Stability of boson stars,'' {\em Physical Review D}, vol.~38,
  no.~8, p.~2376, 1988.

\bibitem{seidel1990dynamical}
E.~Seidel and W.-M. Suen, ``Dynamical evolution of boson stars: Perturbing the
  ground state,'' {\em Physical Review D}, vol.~42, no.~2, p.~384, 1990.

\bibitem{kusmartsev1991gravitational}
F.~V. Kusmartsev, E.~W. Mielke, and F.~E. Schunck, ``Gravitational stability of
  boson stars,'' {\em Physical Review D}, vol.~43, no.~12, p.~3895, 1991.

\bibitem{brito2023stability}
M.~Brito, C.~Herdeiro, E.~Radu, N.~Sanchis-Gual, and M.~Zilh{\~a}o, ``Stability
  and physical properties of spherical excited scalar boson stars,'' {\em
  Physical Review D}, vol.~107, no.~8, p.~084022, 2023.

\bibitem{siemonsen2021stability}
N.~Siemonsen and W.~E. East, ``Stability of rotating scalar boson stars with
  nonlinear interactions,'' {\em Physical Review D}, vol.~103, no.~4,
  p.~044022, 2021.

\bibitem{lee1992nontopological}
T.-D. Lee and Y.~Pang, ``Nontopological solitons,'' {\em Physics Reports},
  vol.~221, no.~5-6, pp.~251--350, 1992.

\bibitem{cardoso2022eco}
V.~Cardoso, C.~F. Macedo, K.-i. Maeda, and H.~Okawa, ``Eco-spotting: looking
  for extremely compact objects with bosonic fields,'' {\em Classical and
  Quantum Gravity}, vol.~39, no.~3, p.~034001, 2022.

\bibitem{kibble1980some}
T.~W. Kibble, ``Some implications of a cosmological phase transition,'' {\em
  Physics Reports}, vol.~67, no.~1, pp.~183--199, 1980.

\bibitem{zurek1985cosmological}
W.~H. Zurek, ``Cosmological experiments in superfluid helium?,'' {\em Nature},
  vol.~317, no.~6037, pp.~505--508, 1985.

\bibitem{ma2024asymmetric}
T.-C. Ma, H.-Q. Shi, and H.-Q. Zhang, ``Asymmetric symmetry breaking: Unequal
  probabilities of vacuum selection,'' {\em arXiv preprint arXiv:2405.05168},
  2024.

\end{thebibliography}

\end{document}